%% file: mgii.tex
\def\etal{{\it et~al.~}}
\def\Lya{{\rm Ly}\kern 0.1em$\alpha$}
\def\Mg{{\rm Mg}\kern 0.1em{\sc II}}
\def\MgII{{\rm Mg}\kern 0.1em{\sc II}~$\lambda\lambda2976, 2803$}
\def\C{{\rm C}\kern 0.1em{\sc IV}}
\def\CIV{C\kern 0.1em{\sc IV}~$\lambda\lambda1548, 1550$}
\def\HI{{\rm H}\kern 0.1em{\sc I}}
\def\kms{\hbox{km~s$^{-1}$}}
\def\cm2{\hbox{cm$^{-2}$}}
\documentstyle[aas2pp4]{article}

\tighten
\begin{document}


\slugcomment{\it submitted to Astrophysical Journal}
\lefthead{Charlton \& Churchill}
\righthead{{\Mg} Absorbing Galaxies}


\title{\LARGE\bf Mg~II Absorbing Galaxies: Halos or Disks?}
\pagestyle{empty}

\author{\Large\sc Jane~C.~Charlton\altaffilmark{1}}
\affil{\normalsize\rm Astronomy and Astrophysics Department \\
       Pennsylvania State University,
       University Park, PA 16802 \\
       email: {\it charlton@astro.psu.edu}}
\author{\Large\sc Christopher~W.~Churchill}
\affil{\normalsize\rm Board of Studies in Astronomy and Astrophysics \\
       University of California, Santa Cruz,
       Santa Cruz, CA 95064 \\ 
       email: {\it cwc@ucolick.org}}

\begin{center}
Accepted for publication: {\it Astrophysical Journal}
\end{center}

\altaffiltext{1}{Center for Gravitational Physics and Geometry,
                 Pennsylvania State University}

\pagestyle{empty}

\begin{abstract}

We challenge the conventional view that the majority of {\Mg}/Lyman
limit absorbers are extended halos of galaxies comprised of
``clouds'' with near--unity covering factor.
The gaseous disks of spiral galaxies are known to extend to large radii
and likely contribute a significant cross--section for absorption.
We perform a Monte--Carlo survey of QSO fields in which the
model galaxies have {\Mg} absorbing ``clouds'' in a spherical
halo or in a randomly oriented disk.
For both geometries, models that recover the 
observed properties of {\Mg} absorbers have only 
a 70--80\% covering factor.
Therefore, regardless of absorber geometry, a
survey of randomly selected QSO fields should
yield a non--negligible number of non--absorbing galaxies
at small impact parameters.  Very few have
been observed (\cite{ste95}).  However, selection effects
are important, and once the observational procedures are 
applied to our model fields we find that this
result is expected.  Since both spherical halo and disk models
can be made consistent with survey results, we present
tests for discerning the geometric
distribution of absorbing gas from {\it Hubble Space Telescope}\/
images and high resolution spectra.

\end{abstract}


\section{Introduction}
\pagestyle{myheadings}
\markboth{\sc Charlton \& Churchill \hfill {\Mg} Absorbing Galaxies~~}
         {\sc Charlton \& Churchill \hfill {\Mg} Absorbing Galaxies~~}

Quasar absorption lines (QALs) provide unique, 
sensitive, and direct probes of the gaseous content in galaxies
over look--back times comparable to a Hubble time.
Studies of QALs promise to contribute insights into the
processes that regulate the evolution of galaxies and the
mechanisms that determine the physical conditions in interstellar and
halo clouds.
The gas in galaxies is a key ingredient to their on--going
evolution, yet virtually nothing is known about the gaseous conditions
in early--epoch galaxies, and how these conditions evolve to the
present time.
The field of QSO absorption lines has matured quite rapidly in the
last half--decade.  
As such, a reasonably sized database, from which we can draw
inferences about the general properties of galaxies known to give rise
to absorption, has accumulated.

Low resolution spectroscopic surveys of the resonant {\CIV} and
{\MgII} doublets (\cite{sbs88,ssb88,ss92}) have established the
statistical nature and evolution of various ``populations'' of
absorbers. 
The evolution of the population selected by the presence of {\Mg} is
virtually indistinguishable from the population selected by the
presence of a Lyman limit break.
This comes as little surprise, since {\Mg} is known to trace {\HI} gas
that is optically thick at the Lyman limit (\cite{ber86,ss92}).
Follow--up imaging surveys of {\Mg} absorbers have been highly
successful at detecting a galaxy at the redshift seen in absorption
(\cite{ber91,ste95}).
The associated galaxies are seldom fainter than 0.1~$L_K^*$
and span the full range of galaxy types (\cite{sdp94,ste95,drink95}).
However, galaxies in the same cluster as the QSO provide an
exception to the rule that {\it every}\/ bright galaxy produces {\Mg}
absorption (\cite{bec92}).
This could occur more generally in galaxy clusters where ram pressure
or high energy ionization fields act to destroy extended low
ionization gaseous components.

It has become apparent that these various populations are likely the
absorption signatures of various regions and conditions within galaxies.
Indeed, the survey data have been used to infer the statistical
properties of a ``typical'' {\Mg} absorbing galaxy at intermediate
redshifts (\cite{ste93a}).  
Such a galaxy is characterized by an inner region of radius $\sim
15$~kpc that gives rise to damped {\Lya} lines, a region extending to
$\sim 40$~kpc that produces {\Mg} absorption and a Lyman limit break,
and an outer region extending to $\sim 70$~kpc that is less
shielded from the extragalactic background radiation and produces
absorption lines in higher ions such as {\C}.  
Beyond these inferences drawn from statistical cross--section arguments,
there has been little direct information on which parts of galaxies
give rise to {\Mg} absorption.
Studies through the Galactic Halo (\cite{sav93,sav95,sem95}) have
provided results which suggest that complex dynamical
disk/halo interactions may give rise to {\Mg} absorbing gas.  
These processes may have analogs in external galaxies, but definitive
interpretations of absorption from the Galaxy are lacking due to our
difficult vantage point.
Bowen, Blades, \& Pettini (1995a,b) examined UV spectra and images
(often in several bands, including X--ray and {\HI}) of
local {\Mg} absorbing galaxies.  
They often found that it is a challenge to unambiguously
identify the substructures giving rise to {\Mg} absorption in these
nearby systems, which often showed some sign of past interaction or
were accompanied by satellite galaxies.
In other words, the environments around the galaxies were not always
isolated enough to clearly determine with which object the gas was
actually associated.
In the case of a line of sight through M81 and the Galaxy, each of
which have satellite galaxies affecting their environments, the {\Mg}
absorption profile is blended with a 400~{\kms} spread.
A further important lesson from their study is the possibility of
altogether mis--identifying the absorbing galaxy.
In the case of Q$1543+489$, two galaxies with velocity separation
120~{\kms} are absorber candidates.
The first has impact parameter 45$h^{-1}$~kpc, and the second
83$h^{-1}$~kpc.  This latter galaxy is outside the range probed
by existing intermediate redshift imaging surveys,
and as such, would be completely missed.
Indeed, such surveys would likely claim they had successfully
identified the former galaxy as the absorber.
Such mis--identifications or ambiguities, even if they are relatively
infrequent, may have an impact on inferences drawn from small number
statistics, such as the fraction of non--absorbing galaxies at small
impact parameters.

Ultimately, the kinematic--rich high resolution absorption spectra
obtainable with 10--meter class telescopes and the high spatial 
resolution Hubble Space Telescope (HST) images should aid
the quest to understand (in individual cases) which types
and parts of galaxies give rise to {\Mg} absorption.
Indeed, high resolution spectra have provided some leverage for
examining statistical correlations between host galaxy properties and
absorbing gas kinematics (also chemical and ionization conditions).
Lanzetta and Bowen (1992) studied five absorbing galaxies at
resolutions of 7 and 35 {\kms} and found some profiles to be 
consistent with the signatures expected from a simple
rotating structure and some that match predictions from
infall/outflow within a spherically symmetric structure.
Additionally, they found preliminary evidence that the profile
complexity and the strength of absorption are correlated with impact
parameter (eg.~only one or two subcomponents at larger impacts).
However, as concluded by Bowen, Blades, \& Pettini (1995b)
from their study of nearby
galaxies (where high spatial resolution data are available and
evidence for interactions has been seen), generalized models of halos
may be misleading.
Detailed models of many individual systems may provide a database
from which a more general understanding of absorbing gas can
be obtained.
Churchill, Vogt, \& Steidel (1995) 
presented a preliminary study of a striking HIRES 
(\cite{vog94}) profile from Q$1331+170$, and suggest that
quasi--symmetric outflowing structures (eg.~superbubbles) may be
present in what appears to be two close galaxies.
Unfortunately, such a model can never be demonstrated to be a unique
interpretation of the data.
Clearly, even with HST images to corroborate inferences based upon
high resolution spectral data, a large sample will
need to be compiled before a comprehensive understanding of the
spatial distribution, motions, and origins of galactic substructures 
giving rise to {\Mg} absorption is developed.

Nonetheless, now that a database of {\Mg} absorbing galaxy properties
(eg.~luminosities, optical/IR colors, redshifts, and impact
parameters) is becoming available (\cite{sdp96}, hereafter SDP),
we can begin to learn something about the
absorbing gas and its relation to the host galaxy. 
Though the high resolution spectra provide detailed information about
the absorbing gas, existing low resolution absorption line data are
proving to be sufficient for developing a basic appreciation of the
conditions under which the presence of {\Mg} is expected.

\subsection{This Paper}

In this paper, we are motivated by the basic question: 
``what is the general {\it geometric}\/ cross--section of {\Mg}
absorbing gas in and around galaxies?''
What we really aim to learn is if a single geometric cross--section
can be invoked to help predict the presence or non--presence of
absorption once a galaxy's general properties and line of sight
impact to the QSO are known.
The conventional wisdom is that {\Mg} absorption is produced in
effectively spherical extended halos [dating from Bahcall \& Spitzer
(1969)].  
Recent support for this view has been presented by Steidel (1995)
(hereafter S95), who has demonstrated that the physical extent of 
{\Mg} absorbing gas around galaxies scales with rest $K$ luminosity 
following a Holmberg--like relationship (\cite{hol75})
\begin{equation}
R (L_K) = 38 h^{-1} \left( L_K/L^*_K\right) ^{0.15}~~\hbox{kpc} 
\end{equation}
(cf.~Fig.~2 in S95).  
The S95 paper presents preliminary results
from the extensive study of SDP which is still in progress.
One striking result to date is that of 58 absorbing galaxies and 14
non--absorbing galaxies measured, there are only two non--absorbing galaxies
``below the $R(L_K)$ line'' given by this relationship (i.e.~only two
galaxies have line of sight impact parameters that fall within the now
predicted surrounding absorbing region and do not exhibit absorption).
Also, in only three cases is the identified absorbing galaxy found 
``above the $R(L_K)$ line''.
The most straight--forward interpretation is that {\Mg} absorbing gas
is distributed with a roughly spherical cross--section, and has a
nearly unity covering factor down to the Steidel \& Sargent (1992)
survey equivalent width limit $W_0(\lambda 2796) \sim 0.3~\hbox{\AA}$ with
a $K$ luminosity (mass) dependent ``cut--off boundary''.
This conclusion is supported in part by the notion that a disk--like
geometry, of random orientation, presents a significantly smaller
statistical cross--section such that a non--negligible number
of non--absorbers below the $R(L_K)$ line would have been detected.

However, these simple geometric models deserve further consideration,
since there is clear evidence for ``clumpiness'' in the absorbing gas. 
Though Lanzetta \& Bowen (1990) found that $W_0(\lambda 2796)$ and the impact
parameter $D$ of the absorbing galaxies follow the ``smooth'' relation
$W_0 \propto D^{-\alpha}$ with $\alpha = 0.92 \pm 0.16$, S95 has shown
that in fact there is a large scatter, which is not consistent with
the tight correlation expected to arise from a smooth distribution of
halo gas (cf.~Fig.~3 of S95).  
Since intermediate and high resolution absorption lines typically
exhibit multiple absorption subcomponents (\cite{pet90,lan92}), we
infer that a near--unity covering factor must result from absorbing
gas distributed in discrete clumps which each have a minimum
equivalent width $W_{\rm min}$ set by the physics of photoionized
diffuse gas.  
We also note that the evidence for unity covering factor is
statistical -- in any single case it is not possible to be certain
that the galaxy identified at the same redshift as that seen in {\Mg}
absorption is indeed the galaxy responsible for the absorption.

We will demonstrate that galaxy disks with realistic aspect ratios
contribute a geometric cross--section for {\Mg} absorption competitive
with that of spherical halos.
We also examine the possibility that observed absorption properties
may actually be governed to a large extent by gas distributed in a
disk geometry.  
In \S3 we describe Monte--Carlo simulations of populations of galaxies
with absorbing gas distributed in spherical and in disk geometries,
and compare these models to the observed impact parameters,
luminosities, and equivalent widths in the SDP survey.
A careful consideration of the selection procedures, given in \S4,
is essential to the interpretation of our model results. 
In \S5, we present simple predictions, to help distinguish whether absorption
often arises primarily in the disks themselves, based upon forth--coming
high spatial resolution HST images of the absorbing 
galaxies and from high resolution spectra of the absorption lines.
The basic results of our Monte--Carlo study of the cross--section
of {\Mg} absorption are summarized in the concluding \S6.


\section{The Statistical Contribution of Disk Geometries}

It is expected that some fraction of the {\Mg} absorbing galaxies
likely result from lines of sight through the disks of Milky Way--like
galaxies.
High velocity clouds in the Halo have a covering factor of at most
38\% [through half the halo (\cite{sav95,bow95a})], and thus by 
themselves do not provide a near--unity covering factor.
Further, Bowen, Blades, \& Pettini (1995a) find that a
significant fraction of the high
velocity clouds are kinematically consistent with disk co--rotation,
so that they all may not really be a halo component {\it per se}, but
instead be more closely associated with the disk.

Since roughly 80\% of field galaxies\footnote{The presence of {\Mg}
absorption does select the population of field galaxies
(\cite{ber91,ste93a}).} have disks, we consider the {\Mg} absorption
cross--section contributed by a randomly oriented population of disks.
In fact, a disk embedded in a sphere of the same radius ($R_d/R_h =
1$) presents a significant cross--section.
An infinitely thin disk yields a lower limit of {\onehalf} for the
relative probability for intersecting a disk relative to a sphere.
When the disk thickness is increased, or the condition $R_d/R_h
= 1$ is relaxed, this probability increases rapidly.
We define the disk thickness $T = h/R_d$, where $h$ is the
half--thickness of the disk, and derive the relative probability for 
intersection of a population of randomly oriented disks with
respect to spheres of radius $R_h$,
\begin{equation}
P(T,R_d/R_h) = (T+0.5) \left( \frac{R_d}{R_h} \right) ^2 .
\end{equation}
As illustrated in Fig.~1, to achieve equal cross--section
contributions from a population of disks and spheres, the absorbing gas
in a disk need only extend $\sim 20$\% beyond that in a spherical
structure for reasonable thickness.
Values of $T=0.2-0.3$ are plausible for the outer, warped extensions
of galaxy disks (\cite{dip91}), since a disk with a scale--height of
5~kpc would likely have a significant amount of {\it absorbing}\/ gas
at twice that height.
Furthermore, it is not unreasonable to
conjecture that {\Mg} absorbing gas in these regions could be as
extended or even more extended than a distribution of {\Mg} absorbing
gas in a halo, given that the {\HI} necessary for photoionization
shielding is likely more concentrated in the plane defining the disk.

It is also important to consider what the effective covering factor, down to
some equivalent width limit, would be for a population of disks.
We suggest that the equivalent width is likely to increase with
disk inclination because the increased pathlength through an inclined
disk yields increased column densities and a larger velocity
dispersion of absorbing clouds, since the number intercepted is
proportional to pathlength.
A simple increase in column density increases $W_0(\lambda 2796)$ most
effectively when absorption falls on the linear part of the curve of
growth.
A curve of growth analysis reveals that, near the equivalent width
limit of 0.3~{\AA}, absorption is on the linear part of the curve
of growth for $b \geq 20$~{\kms}.
Doppler parameters on this order are quite common in the intermediate
resolution ($\sim 30$~{\kms}) spectra of Petitjean \& Bergeron (1990), 
and are consistent with the idea that low resolution data sample the velocity
dispersion of a few or more individual clouds along the line of sight.
For $W_0(\lambda 2796) \geq 0.6$~{\AA}, it is even more likely
that the number of clouds intercepted and their velocity dispersion
dominates the value of the equivalent width.
Petitjean \& Bergeron (1990) find that $W_0(\lambda 2796)$ still 
scales linearly with the number of resolved subcomponents.

Ultimately, the extent of the disk contribution to the overall 
cross--section for {\Mg} absorption depends upon the detailed physics
of sharp {\HI} edges (gas dynamical processes, ionization field and
strength, galaxy mass and shape, external pressure)
(\cite{mal93,cor93,dov94}).
Locally, extended {\HI} disks (beyond the optical radius) are common
[cf.~the recent review by Irwin (1995) and references therein], but
only in a few cases has the $N(${\HI}$)$ distribution been mapped down
to a sensitivity below the expected sharp edge at which the gas
converts from mostly neutral to mostly ionized (\cite{cor89,van93}).  
In NGC~3198, the $N(${\HI}$)$ distribution is observed to drop sharply with
radius for $N(${\HI}$) < 5 \times 10^{19}$~{\cm2} (\cite{van93}),
but is still $ 10^{18}$~{\cm2} at 35--40~kpc.
This rapid decline is likely to slow at larger radii.  
Bowen, Blades, \& Pettini (1995b) find that there exists 
a correlation between the 
$B$ magnitude of spiral galaxies and their {\HI} radii measured from
21--cm emission.  
Their application of the Holmberg--like relation $R(${\HI}$) =
R^* (L_B/L^*_B)^{\beta}$ to local {\HI} disks provides evidence that
$\beta$ becomes less steep as the $N(${\HI}$)$ threshold is decreased, and
approaches the values required to match the intermediate redshift
{\Mg} absorbers, namely $\beta \sim 0.2$ [for $B$ luminosity (S95)] at
the Lyman limit, $N(${\HI}$) \sim 10^{17.3}$~{\cm2}.
Although it may be an extreme example, we also note that in the M81
interacting group there is a large covering factor of material with
$N(${\HI}$) > 10^{20}$~{\cm2} beyond a radius of 40~kpc
around M81 itself, which is spread throughout the group in a flattened
distribution (\cite{yun94}).

In the Milky Way, $N(${\HI}$) > 10^{19}$~{\cm2} at a galactocentric
radius of 30~kpc in the direction $l=130^{\circ}$ (\cite{dip91}).  
By extrapolating this value using their observed radial
scale--length of 5.5~kpc in this direction, we obtain
$N(${\HI}$) = 10^{17}$~{\cm2} at a radius of $\sim 60$~kpc.
HST spectra of various QSOs, Active Galactic Nuclei, and Halo
stars, looking in various directions through the Galaxy, show
{\Mg} absorption arising in the disk (\cite{sav93,lu94,car95,sem95}),
though the radial extent of the {\Mg} itself is not observed directly.
In some cases, the absorbing gas could actually be associated with an
$N(${\HI}$)$ even smaller than $10^{17}$~{\cm2} [at high redshift {\Mg}
is found to be associated only with systems that are optically thick
at the Lyman limit, but at low redshift the ionizing background is
reduced and some {\Mg} systems are observed with $\tau_{LL}$ somewhat
less than unity (\cite{ber94})].
Thus, in the Milky Way, it might even be expected that the
{\Mg} disk extends well beyond 50~kpc.
To make all this plausible, metals would have to exist at these
large radii, either by the action of galactic fountains, by
infall of debris from satellites, or because they were present in
the material from which the galaxy was formed.

Based upon these arguments, we suggest that
the cross--section contribution of thick and/or extended disks and
their related material (eg.~co--rotating clouds) may well be large
enough to account for the majority of {\Mg} absorbers.
Realistically, absorbers are probably quite diverse, including 
contributions from LMC--like satellites, streaming tidal debris formed
in merging events, occasional $\tau_{LL}$ low surface brightness
galaxies, and even early--type galaxies of sufficient $K$ luminosity
[the latter two having been observed in the SDP survey (Steidel,
private communication)].


\section{Modeling Absorbing Galaxies}

We perform Monte--Carlo simulations that allow us to predict the
numbers and properties of absorbing and non--absorbing galaxies in a
sample of randomly generated fields centered on QSOs.
We have designed the simulations so that our results can be
compared to observations.
Since the SDP survey represents the most comprehensive observational program to
date, we briefly describe their work [but also see Drinkwater, Webster,
and Thomas (1995)].

Drawing primarily from the Sargent, Steidel, \& Boksenberg (1988)
and Steidel \& Sargent (1992) surveys, SDP
obtained optical and IR images of 51 QSO fields in which 58 {\Mg}
absorbers had been detected ($0.3 \leq z \leq 0.9$)
and of 25 ``control'' fields in which the QSO spectra exhibit no
absorption to a completeness limit of $W_0 (\lambda 2796) =
0.3$~{\AA}.  
The images were taken over a large range of band passes and the
detection levels were targeted to $\sim 2$ magnitudes fainter than present
day $L^*$ for the known absorption redshifts.
A combination of filters were used in order that the rest--frame $B$
magnitude of the galaxies could be measured.
Additionally, $K$ band images were obtained for all fields.
Confirmation of any absorbing galaxies identified in the imaging
portion of the program required follow--up spectroscopy.
Unless a field had several identified candidates, spectra were
obtained in a pattern from small angular separation to larger until a
match was found for the redshift seen in absorption. 
This follow--up spectroscopy in most fields is nearly complete within
a $8-10{\arcsec}$ radius of the QSO.  For control fields, redshifts
were obtained for nearly all galaxies in the images within 
$10{\arcsec}$ of the QSO.  Preliminary results from this study
are summarized in S95.

For our model fields, the redshifts, luminosities, and impact
parameters are known for {\it all}\/ galaxies within $10\arcsec$ of the
field center.  The fraction of absorbing and non--absorbing model galaxies
is thus representative of an unbiased sample of QSO fields.
Our procedure is as follows:
The galaxy redshift is selected at random in the range $0.3 \leq z
\leq 0.9$, roughly matching the observed distribution.
The redshift determines the physical size of the region (corresponding
to a radius of 10{\arcsec}) within which this galaxy will be
located\footnote{The conversion assumes $q_0 = 0$.  All sizes quoted
throughout this paper would be $\sim 80$\% smaller for $q_0 = 0.5$.}.
The galaxy luminosity $L_K$ is selected from a Schechter function,
\begin{equation}
P(L_K)dL_K \sim (L_K/L^*_K)^{\alpha} \exp \left[-(L_K/L^*_K) \right] dL_K,
\end{equation}
with $\alpha = -1.0$.
This is derived from the value, $\alpha = -0.7$, observed
for the luminosity function of the absorption--selected sample (S95), 
which is then weighted inversely by the absorbing cross--section area
of a galaxy to obtain the true luminosity function.
The galaxy is located randomly within the 10{\arcsec} region 
surrounding the QSO, giving the impact parameter $D$.
The ``observed'' equivalent width $W_0$ along the line of sight is then
determined based on the assumed geometry and model parameters.  
If the generated $W_0$ exceeds the 0.3~{\AA} observation threshold,
then the galaxy is designated as an ``absorber'', and otherwise as a
``non--absorber''.

\subsection{Spherical Models}

Invoking a spherical geometry, S95 suggested
that the relatively sharp observed cut--off in the size of
{\Mg} absorbing regions may be provided by the decrease of the
pressure of diffuse gas that confines the clouds responsible for
{\Mg} absorption within the halo.
When the pressure decreases below a threshold value, which scales with
the mass of the galaxy, the column density of {\HI} in an individual
cloud is no longer sufficient to shield {\Mg} and keep it in its
singly ionized state.  
This effect roughly sets a minimum column density for {\it individual
clouds}, for as the pressure decreases outward from the galaxy center,
$N(${\HI}$)$ of the clouds decreases (\cite{mo95}).
This column density limit apparently corresponds to an equivalent
width of $0.15-0.30$~{\AA}, since weaker absorbers are rarely observed
(\cite{ss92,chu96}).
To investigate this supposition, we ran a series of photoionization
models, CLOUDY (\cite{ferland88}), and computed the equivalent widths
of model clouds.
These models were plane parallel slabs illuminated from one side with
continuum slope $\alpha_{\rm ox} = -1.5$.
For the range of ionization parameters $-4.0 \leq \log \Gamma \leq
-1.0$, the average equivalent width was 0.13~{\AA} for $\tau_{LL} \sim
1$ clouds with $\log N(${\HI}$) \sim 17.5$~{\cm2}.  

Empirically, the distribution of equivalent widths
is roughly consistent with a power law (\cite{ss92}),
\begin{equation}
n(W_0)dW_0 \sim W_0^{-\delta}dW_0 , 
\end{equation}
where $\delta = 1.65 \pm 0.90$, and thus many absorption
systems have total $W_0$ near this limiting equivalent width.
The relationship between the sum of the column densities and the
observed equivalent width depends upon the distribution of velocities
of clouds along the line of sight. 
Intermediate resolution spectroscopy reveals that the equivalent width
increases almost linearly with the number of components
(\cite{pet90}).  
Thus, it is reasonable to assume that for the low resolution spectra,
the individual clouds are sufficiently dispersed in velocity space
so that their equivalent widths add to a total observed equivalent
width $W_0$ for the {\Mg} absorber.

The challenge for designing realistic spherical cloud models is to
produce a near--unity covering factor when a typical line of sight at
large impact is likely to pass through only one or two clouds.
If this is the case, then some lines of sight will certainly fail
to pass through any clouds observable in {\Mg}, resulting in
non--absorbers.

\subsubsection{Design and Parameters}

Guided by this general picture, we design models of a distribution of
clouds in a spherical halo, governed by the following parameters: 
(1)~a mean number of clouds $N_{\rm c}(0)$ along a line of sight
through the halo with impact parameter $D=0$,
(2)~a power law dependence of the equivalent width of an individual
cloud on the radius from sphere center, $W(R) \sim R^{-\alpha}$,
that would result from a decrease in the individual
cloud column densities due to decreasing pressure with increasing $R$,
(3)~a variable cloud size factor, given as the variance $\sigma$
of a Gaussian distribution centered on $W(R)$, and
(4)~a mean minimum limiting equivalent width $W_{\rm min}$ for an
individual cloud.

For a line of sight at impact parameter $D$, the number of clouds
$N_{\rm c}(D)$ is determined by scaling $N_{\rm c}(0)$ by the
pathlength through a sphere of radius $R(L_K)$ [given by Eq.~(1)].
In each model we choose either $W_{\rm min} =$ 0.15~{\AA} or 0.30~{\AA}.
For the former, the mean number of clouds at $R(L_K)$ is set
equal to two, and for the latter to one, so that the mean
$W_0$ at $R(L_K)$ is 0.30~{\AA} in either case.
The value of $R(L_K)$ is determined using $\beta=0.15$ 
to match observations (S95),
however, we adjust the normalization $R^*$ to minimize the sum of the
number of non--absorbers below and the number of absorbers above the
$R(L_K)$ line given by Eq.~(1).  
This iterative process results in self--consistent models.
The actual number of clouds along the particular line of sight is
selected from a Poisson distribution with
mean $N_{\rm c}(D)$.  
The locations of the clouds are chosen at random along the line of
sight (no dependence of cloud density on $R$), and the power law index 
$\alpha$ is used to compute $W(R)$.
The actual individual cloud equivalent width $W_{\rm c}$ is selected
from a Gaussian distribution with variance $\sigma$ centered on
$W(R)$.
The ``observed'' equivalent width $W_0$ is determined by summing the
$W_{\rm c}$, an approximation justified by the nearly one to one
correlation seen between equivalent width and the number of
subcomponents in an absorption line (\cite{pet90}).
A successful model should recover the observed equivalent width
distribution $n(W_0)dW_0$.
In order to roughly constrain our models, we assume a power law
distribution and determine the least squares fit slope $\delta$
to the distribution of model $W_0$.

\subsubsection{Absorption Properties}

Table 1 lists the various series of model parameters and 
the fractions of absorbers above $f_{\rm above}$ and below $f_{\rm
below}$ the $R(L_K)$ boundary [shown in Fig.~2 and given by
Eq.~(1)] for an unbiased sample of fields.
In fact, the fraction $f_{\rm below}$ gives the covering factor
within radius $R(L_K)$.
From models S1--S4, we see that $\delta$ is strongly dependent on the
value of $\alpha$, but that $\alpha$ has little effect on the covering
factor.  
There is a strong dependence of $\delta$ on the mean number of clouds
along a line of sight (models S5--S7).
The dependence of these gross properties on $\sigma$ is minimal
(compare models S3 and S6).
By increasing the number of clouds $N_{\rm c}(0)$, the covering 
factor $f_{\rm below}$
can be increased such that it is quite close to unity, and $\alpha$ can be
chosen so that the observed value of $\delta \sim 1.65$ is obtained.
In these spherical cloud models, large $W_0$ result from multiple clouds
along a line or sight and/or from a large value of $\alpha$.
Table 1 illustrates the balance between the parameters $N_{\rm c}(0)$ 
and $\alpha$.
For $\alpha = 2$, we can obtain $\delta = 1.6$ (enough large $W_0$) with
a mean of only two clouds at $D=0$ (model S11).  
For $\alpha = 0.5$, a mean of three clouds is necessary (model S2).

The covering factor could be further increased by using an even
larger $N_{\rm c}(0)$, but then there would be too many large $W_0$ values
($\delta$ would be too small for any $\alpha > 0$).
This problem is less severe if $W_{\rm min}$ is reduced from 0.30~{\AA}
to 0.15~{\AA} (compare models S7 and S10), but the lack of observed
small $W_{\rm min}$ values, discussed above, indicates that 0.15~{\AA}
is really a limit on how far down this parameter can be pushed.
For models S2, S11, and S12, we set $N_{\rm c} (0)=3$, $2$, and $6$,
and tuned $W_{\rm min}$, $\alpha$ and $\sigma$ to match the
observed $n(W_0)dW_0$ distribution.
These three models will be used in the following sections to
illustrate variations between different realizations of spherical
cloud models.

The predicted distribution of impact parameter $D$ versus $K$
luminosity is illustrated in Figs.~2(a--c) for three different
randomly selected realizations of model S2.
To facilitate some level of comparison to Fig.~2 of S95, we plot 58
random absorbing galaxies.
The number of non--absorbers (63) presented on this figure is the mean
number that would arise in the same number of fields that yield a mean
of 58 absorbers.
At first glance, the lack of a clean demarcation of absorbers and 
non--absorbers around the $R(L_K)$ line appears inconsistent with the
SDP data.  As we will address in \S4, the appearance and subsequent
interpretation of a $D$ versus $L_K$ diagram may be quite
sensitive to observational selection procedures.  In fact, models
S2 and S12 succeed fairly well in that they yield a large value
of $f_{\rm below}$ and a small value of $f_{\rm above}$.
However, we note that this is not the case for all spherical
cloud models that we can design.  For example, Table 1 shows that
in model S11 a much larger fraction ($0.34$) of the galaxies below
the $R(L_K)$ line will be non--absorbers and a much larger fraction
($0.22$) of the galaxies above the $R(L_K)$ line will be absorbers.
In these models, absorbing galaxies can exist above the $R(L_K)$ line
because of the Poisson fluctuations in the number of clouds
along the line of sight and the Gaussian spread around the
equivalent width of a single cloud.  
From Table 1, we infer that the parameter
$N_{\rm c} (0)$ is the most critical in determining the value of
$f_{\rm above}$.  For $N_{\rm c} (0) = 2$ the value of
$N_{\rm c} (D)$ will decrease slowly with $D$ and the probability of
detecting one cloud can be significant for an impact greater than $D =
R(L_K)$.

In Figs.~3(a--c), we also illustrate the distributions of $W_0$ versus
$D$ for the three realizations presented in Figs.~2(a--c).
These diagrams illustrate that our approach to modeling ``clumpiness''
within the absorbers manifests in a $W_0 - D$ distribution qualitatively
consistent with the data presented in Fig.~3 of S95.
The scatter in the diagram at a given $D$ value is produced by
the variation of the number of clouds along a line of sight, and
the various locations (in galactocentric distance $R$) at which clouds
are intercepted.

\subsection{Disk Models}

Expected properties of absorbing gas in galaxy disks should be
guided by impressions of the thick, warped outer disk regions
of nearby spirals, as we have discussed in \S2.  
There are likely to be clumps and irregularities (``clouds'') in the disk
(\cite{irw95}, and references therein).
The number of clouds and their overall velocity dispersion is likely
to increase with increased pathlength through an inclined disk.
Based upon the conclusions of Bowen, Blades, \& Pettini (1995b)
that the extent of {\Mg} absorbing gas in
disks is likely follow the Holmberg--like relation for $B$ luminosity with
$\beta \sim 0.2$, we assume that absorbing gas in disks roughly follows
Eq.~(1) for $K$ luminosity with $\beta = 0.15$.

\subsubsection{Design and Parameters}

The parameters for our disk models are:
(1)~a power law dependence of $W(R)$ on the radius in the disk,
given by 
\begin{equation}
W(R) = 0.3 \left( R/R_{0.3} \right) ^{-\alpha}~~\hbox{\AA} ,
\end{equation}
where $R_{0.3}$ is the radius at which a line of sight perpendicular
to the disk would yield $W_0 = 0.3$~{\AA} on average,
(2)~a ``clumpiness'' factor, given as the variance $\sigma$ of a
Gaussian distribution centered on $W(R)$, and 
(3) a ``cut--off radius'' $R_{\rm cut}$ beyond which $W(R)$
effectively vanishes, which could be somewhat larger than $R_{0.3}$ for
the less severe ionization conditions likely at low redshifts
(\cite{ber94}).

In order to determine $W_0$ for a given line of sight, the disk
is randomly oriented with inclination $i$ [with probability $P(i)di
\sim \sin (i)di$].
The line of sight is located at a random angle $\theta$ from the
major axis of the ellipse representing the inclined disk projected
onto the plane of the sky.  
The radial position where the line of sight pierces the plane of the
disk is then
\begin{equation}
R = D \left[ \cos^2{\theta} + 
   \left ( {{\sin{\theta}} \over {\cos{i}}} \right) ^2
     \right]^{1/2}.
\end{equation}
The equivalent width $W(R)$ is given by Eq.~(5) scaled by the
increased pathlength factor $\sec(i)$ to roughly account for the
combined effects of higher line of sight column density and velocity
dispersion of the absorbing clouds.
The ``observed'' $W_0$ is then selected from a Gaussian distribution
with variance $\sigma$ centered on $W(R)$.
From physical principles, we expect a sharp cut--off at some radius in the 
{\Mg} absorbing gas, so we assign a cut--off radius in the range $1.5
\leq R_{\rm cut}/R_{0.3} \leq 2$ (except for one model with $R_{\rm
cut} = \infty$, chosen for purposes of illustration).
As with the spherical models, we normalize Eq.~(1) with $\beta = 0.15$
by minimizing the sum of the number of non--absorbers below
and the number of absorbers above the $R(L_K)$ line.

\subsubsection{Absorption Properties}

In Table 2, we illustrate the effect of varying the disk model parameters.
Note that for a population of infinite disks (model D1), for which
we fix $R^* = 38 h^{-1}$~kpc, there are no non--absorbing galaxies
expected below the $R(L_K)$ line, but clearly disks must have some cut--off.
As $R_{\rm cut}$ is reduced (models D1--D3), the fraction of non--absorbing
galaxies below the $R(L_K)$ line increases, but the sensitivity in the regime
$R_{\rm cut}/R_{0.3} \sim 2$ is small.
From models D4--D6, we note that the fraction of non--absorbers below
the $R(L_K)$ line is not very sensitive to the choice of
$\alpha$.
An inhomogeneous population with $1.5 \leq R_{\rm cut}/R_{0.3} \leq 2$ 
and $1.0 \leq \alpha \leq 2.0$ would yield a similar fraction of non--absorbers
$f_{\rm below} \sim$~0.70--0.75, provided that all galaxies can
be characterized by $R(L_K)=R^*(L_K/L^*_K)^{0.15}$.
Again, a realistic single population model must be consistent with the 
observed distribution $n(W_0)dW_0$ with power $\delta \simeq 1.65$.
Models D7--D9 illustrate the level at which $\alpha$, $\sigma$, and
$R_{\rm cut}$ can be adjusted while still successfully recovering the 
$n(W_0)dW_0$ distribution and showing how the absorption properties vary
between these models.
We select these models for further discussion in the following sections.

Using the same criterion as for spheres, we present the distributions
of $D$ versus $L_K$ for three realizations selected from model D8 (58
absorbers and 58 non--absorbers) in Figs.~2(d--f).
The fraction of galaxies above the $R(L_K)$ line that produce
absorption is larger than $f_{\rm above}=0.15$ for all cases in Table
2.
If $R_{\rm cut}$ is reduced further (below $1.5$), $f_{\rm above}$ can
be reduced, but then the expected fraction
of non--absorbing galaxies below the $R(L_K)$ becomes much larger.
It is difficult to design a realistic disk model that will not produce
a non--negligible fraction of absorbing galaxies above the $R(L_K)$
line.  
Physically, this is because of lines of sight that pass through the
disk at large impact parameters, but also at large inclinations. 

We also present the $W_0$ versus $D$ distribution for 58 absorbers from model
D8 in Figs.~3(d--f).
As with the spherical models, the $W_0$ versus $D$ distribution
is qualitatively consistent with the S95 data, except that
our disk models yield a deficiency of small $W_0$ at small $D$.
This is an artifact of our model design, which does account for
increased pathlength through highly inclined disks using a geometric
scaling, but does not account for material {\it physically}\/
distributed above or below the disk plane (eg.~thick disk or warp
material).  In a more realistic model,
lines of sight at small impacts that graze highly inclined galaxies
with some absorbing gas distributed at $\sim 5-10$~kpc about the disk plane 
[as is likely for the Galaxy (\cite{dip91})] would also exhibit
absorption, filling in the lower left corner of the
$W_0 - D$ distribution. 
It is interesting to note that in S95 most (if not all) of the 
absorbers with $D \leq 15h^{-1}$~kpc are Damped {\Lya} (DLA) absorbers.
This may imply that most DLA absorbers are highly inclined
disks, and that, in view of our model interpretations, extended
material above and below the disk is further implied by their 
detection.

\subsection{Ruminations on an Unbiased Sample}

The bottom line is that, regardless of the overall geometric
cross--section, it is difficult to design models of
absorbing gas which
(1)~consist of physically distinct
clouds as opposed to a smooth distribution of gas that would
exhibit a tighter $W_0 - D$ relationship than is observed,
(2)~recover the distribution of equivalent widths $n(W_0)dW_0$, and
(3)~yield covering factors near unity down to the required equivalent
width limit within some well defined luminosity dependent
galactocentric radius.

We note that none of our spherical cloud models can yield few
non--absorbing galaxies at impact parameters $D \leq R(L_K)$, qualitatively
reproduce the observed $W_0 - D$ distribution, and be made consistent
with the observed equivalent width distribution.
The basic problem is that gaps in covering result from instances
of zero clouds along the line of sight.
This could be adjusted by increasing the number of clouds,
but then the equivalent width distribution is skewed toward large
values.  Decreasing the minimum equivalent width of an individual
cloud could help with this discrepancy, but this would not be 
consistent with the small observed number of absorbers with
$W_0 < 0.3$~\AA.
For disks, we note that taking into account the increased pathlength for
inclined disks does in fact reduce the number of non--absorbers at small
impact parameters.
Indeed, our disk absorber models yield similar fractions of
non--absorbers below the $R(L_K)$ line as do our spherical
absorber models.
In the S95 sample, only three absorbing galaxies are observed above
the $R(L_K)$ line.
Although some spherical cloud models appear to be relatively
consistent with this result, all of our disk models predict larger
numbers of absorbers at large impact parameters.

We must address the fact that our models yield no definitive boundary
between absorbing and non--absorbing galaxies in the $R(L_K) - L_K$ plane.
It is important to take into account that the results
we have discussed so far apply to an ``unbiased sample'' of QSO fields.
For our unbiased sample, there does appear to be a difference between
the predictions for spherical and disk models.
It is {\it in principle}\/ possible to design a spherical
cloud model that predicts very few absorbing galaxies above
the $R(L_K)$ line.  Whether such a model is practically realistic,
effectively whether there is a sharp cut--off in radius for
the existence of absorbing clouds, depends upon the details of 
ionization physics and on the variations in the
confining pressure and cloud properties with radius.
However, none of our disk models produce
a negligible fraction of absorbing galaxies above the
$R(L_K)$ line, and we argue that all disk models would
produce some absorbing galaxies at large impact parameters.

In a unbiased sample of QSO fields (chosen without regard
to the presence or non--presence of absorption) in which all galaxy
redshifts and luminosities are measured to some limiting rest $L_K$,
our ``clumpy'' models predict that there is no definitive boundary
between absorbing and non--absorbing galaxies.
We will show in \S4, that a careful ``application'' of the SDP selection
procedures to our model fields brings our results, for both spherical
and disk models, into plausible agreement with observations.


\section{Confronting Observations}

The SDP sample is not an unbiased sample of QSO fields,
but includes 51 ``absorber fields'' that exhibit {\Mg}
in the redshift range $0.3 \leq z \leq 0.9$ and 25 ``control fields'', in
which absorption is not detected.
In order to better understand the effects of biasing in the sample, 
we studied Monte--Carlo {\it fields}\/ of spherical models S2, S11, and S12
and of disk models D7, D8, and D9.  
An important (and not all that certain) parameter is the mean number of
galaxies per field (within the redshift range).
$N_{\rm field}$ can be estimated from the number of {\it absorbers}\/ observed
per unit redshift, roughly $dN/dz = 0.85 \pm 0.2$ (\cite{ss92}) in the
range $0.3 < z < 0.9$, which translates to $N_{\rm los} = 0.51 \pm 0.12$
absorbing galaxies per line of sight.
From Tables 1 and 2, we find that our simulations predict that the
fraction of
absorbing galaxies per line of sight is $0.46 \leq f_{\rm abs} < 0.54$. 
Thus, the predicted average number of galaxies per field is 
$N_{\rm field} = 0.51/f_{\rm abs} = 0.95-1.10$.
We adopt $N_{\rm field} = 0.95$, but note that the observational
constraint on $N_{\rm los}$ allows some flexibility in this quantity.

We generated 10000 simulated SDP surveys, each having 51 absorber
fields and 25 control fields.  
The galaxies themselves were randomly selected from the models
discussed in \S3 (including redshifts).
Based upon the presence or non--presence of absorption above the 
0.3~{\AA} survey limit, we designated a field as an ``absorber'' or
``control'' field.  
Predictions based upon both the spherical and disk models are quite
similar, and we thus discuss the simulation results more generally.
For $N_{\rm field} = 0.95$ we find that 38--41\% of the fields
($F_{\rm abs} = 0.36 - 0.40$) contain at least one absorbing galaxy.
This is a much smaller fraction of absorbing fields than selected in the
SDP sample ($51/76=0.67$), which may be telling us something about our
models or about a selection bias in the SDP survey.
Simulated absorber fields contain on average 1.24--1.28 absorbers
and 0.44--0.51 non--absorbers [control fields contain
the same number of non--absorbers (0.44--0.51), as they should
since the presence of an absorber does not affect the probability of
finding other galaxies in the field at different redshifts].
On average, the six geometric models yield 63--65 absorbers and 33--39
non--absorbers.
The observed number of non--absorbers (14) is not consistent
(smaller) with the predictions of our models, and the number of
absorbers observed (58) is somewhat improbable, on the low end of the
tail of the distribution.  
In Fig.~4(a), we illustrate the simulated distributions of the number
of absorbers and non--absorbers for model D8 in an ``SDP survey'',
in which all galaxies within a 10{\arcsec} field have been included.
In this figure, we also present the number of non--absorbers with
impact parameters less than 8{\arcsec} which lie below the $R(L_K)$
line.
For this sub--sample, the predicted number non--absorbers approaches
the observed number, providing a powerful illustration of how sensitive
observational results are to observational completeness and selection
effects.

\subsection{Effects of Covering Factor and Selection Bias}

It turns out that the numbers of absorbers, non--absorbers, and the
fraction of absorbing fields $F_{\rm abs}$ are sensitive to the
parameter $N_{\rm field}$.
In Table 3, we illustrate the effect of decreasing $N_{\rm field}$
for models D8 and S12 (other models produce similar numbers).
As illustrated in Fig.~4(b) for model D8 and $N_{\rm field}=0.65$,
the 51 model fields yield a mean of $N_{\rm abs}=59-61$ absorbers,
consistent with the 58 observed by SDP.

A value of $N_{\rm field} =0.65$ is not implausible.
If the observed $dN/dz$ is about 1--$\sigma$ smaller, we recover
$N_{\rm field} =0.65$.
Alternatively, the estimate of $N_{\rm field}$ scales inversely with
the fraction of galaxies in the fields that produce absorption, $f_{\rm
abs}$.
If we had designed models that yield nearly unit covering
factor, we would have increased $f_{\rm abs}$ and provided a means to
reduce $N_{\rm field}$.
However, this would yield an increased number of absorbing galaxies
observed in 51 absorbing fields. 
{\it Galaxy models that yield higher covering factors cannot be made
consistent with the number of absorbing galaxies observed by SDP}.\/
Provided our models statistically represent the absorption properties
of real galaxies, $F_{\rm abs}$ of an unbiased sample of fields may be
as small as 0.28, and the SDP sample may have selected heavily in
favor of absorbing fields.

However, even with $N_{\rm field} \sim 0.65$, our models predict that
the observed number of 14 non--absorbing galaxies is somewhat improbable.
Since the observational procedures of SDP would often miss galaxies at
large angular separations, it is
more instructive to focus on the fact that SDP find only two
non--absorbing galaxies below the $R(L_K)$ line separating absorbers
and non--absorbers.
We will show that this observational result can be statistically
reconciled with our simulations for both spherical and disk models
once we account for
(1) biasing in the choice of fields and in the practical procedure
employed by SDP to obtain follow--up redshifts of galaxies, and (2)
the uncertainty that even a small fraction of the absorbing galaxies
may have been mis--identified (see \S1).

\subsection{Reconciling Selection Effects}

We turn our attention to the fact that a fair number of the non--absorbing
galaxies in the SDP sample do not have measured redshifts, and are
therefore not represented in Fig.~2 of S95.
In practice, the follow--up redshift measurements of galaxies
identified in the SDP fields were implemented using the following
practical considerations (Steidel, private communication).
In control fields, redshifts were obtained for all galaxies within
the $10\arcsec$ fields.
However, in absorber fields, non--absorbers with large impacts are
preferentially missing from the sample, since the search was sometimes
stopped at $\sim 8\arcsec$ once a galaxy with the redshift seen in
absorption had been found.
In nearly all cases, redshifts were obtained for all galaxies within
$8\arcsec$.  

In the context of these procedures, we ask the relevant question -- 
``for our models, what is the probability of detecting only a few
non--absorbing galaxies below a statistically determined $R(L_K)$ line
in a $D$ verses $L_K$ diagram?''.
In order to address the partial coverage in the $8\arcsec \leq D \leq
10\arcsec$ zone, we bracket the SDP procedure with two ``observational
scenarios'': 
(1) all non--absorbers with 10{\arcsec} of the QSO are included, and 
(2) only those non--absorbers within 8{\arcsec} of the QSO are included.
In Table 4, we present the predicted mean number of non--absorbers below the
$R(L_K)$ line based upon these two scenarios for the various
models and for the $N_{\rm field}$ values 0.95 and 0.65.  
From the $N_{10\arcsec}$ and $N_{8\arcsec}$ entries, note that there
is not much sensitivity to the chosen observational scenario.
Nonetheless, in order to best emulate the observational procedures, we
applied scenario (1) to the 25 control fields and scenario (2) to the 51
absorber fields.
We then computed $P(\leq 2)$, the probabilities of observing two 
or fewer absorbers below the $R(L_K)$ line.
We find that both sphere and disk models give non--negligible
probabilities for observing two or fewer non--absorbers in the SDP sample.
There is a 31\% chance (for spherical model S12) that SDP would
detect two or fewer non--absorbers below the $R(L_K)$ line and a 5\%
chance for disk model D8.

Since the SDP observational procedure was to search outwards
in the absorber field until a galaxy was located at the redshift
seen in absorption, it is possible that some ``absorbing'' galaxies
have actually been mis--identified [though, for several reasons, it is
believed that few mis--identifications have been made (S95)].
It is likely that for each mis--identified galaxy, a non--absorbing
galaxy below the $R(L_K)$ line and an absorbing galaxy above the
$R(L_K)$ line would not be represented in the survey.
Thus, in addition to $P(\leq 2)$), we have tabulated $P(\leq 4)$
to examine the effects of possible mis--identifications.
The probability of observing four or fewer non--absorbers is 
not negligible for models S12 (71\%), S2 (40\%), and D8 (26\%), for
$N_{\rm field} = 0.65$ (corresponding to $dN/dz = 0.54$).
This is further illustrated by the distribution function for model D8
presented in Fig.~4.  
The probability of detecting four or fewer non--absorbers is
not negligible in all three disk models, ranging from 16--26\%.

We briefly consider the number of absorbing galaxies above the $R(L_K)$
line.
In Table 5, we present the predicted numbers of absorbing galaxies
above the $R(L_K)$ line and the probabilities of observing fewer than
four such galaxies in an SDP sample.  
Both observational scenarios, eg.~including all
galaxies out to the field limit of 10{\arcsec} or including only those
within 8{\arcsec}, are presented.
In particular, model S12, which is characterized by a large $N_{\rm
c}(0)$ and a small $W_{\rm min}$, results in a large $P(\le 4)$.
As such, it yields the properties of a ``sharp boundary''.
For disk models, there are an abundance of absorbers above the
$R(L_K)$ line in a sample of 10{\arcsec} fields.
For $N_{\rm field} = 0.65$, model D8 predicts a mean of 11 absorbing
galaxies above the line, but only 5 of these are within 8{\arcsec} of
the center of the field and thus certain to be included in the
observational sample. 
If the observational procedures have lead to a bias against detecting
absorbing galaxies at 8--10{\arcsec} then all models have a
non--negligible $P(\le 4)$.

\subsection{Discussion}

Based upon our models and a study of the covering factor issues
implied by the Holmberg--like relationship for absorbing gas
cross--section, we cannot distinguish whether {\Mg} absorption
arises primarily within clouds distributed in a spherical halo or
distributed in a flattened thick disk.
Both of these geometric models appear plausible with the existing
data once we include uncertainties due to possible mis--identifications
and a possible bias in field selection of too few non--absorber
fields.
The number of non--absorbing galaxies observed at small impact
parameters depends strongly on how far a sample of QSO fields is
from an unbiased sample.  
We conclude that the small observed number of non--absorbers in the
SDP sample does not {\it necessarily}\/ imply unity covering factor.
Once observational biases and uncertainties are taken into account, the
covering factor can be made consistent with 70--80\%.
Indeed, if the covering factor does approach unity, it is difficult to
reconcile the {\it small}\/ number of {\it absorbing}\/ galaxies SDP
observed in their 51 fields with models that match all other
observational constraints.

It is clear to us that our models are idealizations of the
universe of real galaxies.
Our disk models do not take into account the fact that some number of
halo clouds are surely present around a realistic disk galaxy.
As a result, disk models yield some non--absorbers for cases
where the line of sight passes too far out in an inclined 
disk to produce absorption (beyond $R_{\rm cut}$), but some of 
these lines of sight might yield absorption in a galaxy with clouds
in an extended halo.
Likewise, our spherical ``halo'' models do not contribute a disk
component, which would increase the number of absorbers.
The implication, if we are to obtain 58 absorbers in 51 fields, is 
that the covering factor of the {\it halo proper}\/ in such model
galaxies would be reduced below $\sim70$\%.
Additionally, our models neglect the presence of satellite galaxies.
Satellite galaxies, merging or otherwise, may contribute a
significant cross--section for absorption (cf.~\cite{york86,wang93}).
From dynamical considerations, it may be expected that absorption
properties may also correlate with the presence of satellites, their
distance from the primary, and the direction of their rotation
with respect to the primary (prograde/retrograde).
Local spiral disk galaxies are known to have an average of one
to two satellites (\cite{dennis92}).  
Interestingly, for their sample, the disk luminosity does not correlate
with the number of satellites, so that absorption due to satellites
would not be expected to correlate with the luminosity of the
primary.
However, they do see the ``Holmberg--effect'', in which the projected
satellites show some evidence for having an excess near the minor
axis of the primary.
Roughly, in the case of the Milky Way, the LMC has $M_B = -18$ (and
$N_H = 10^{19} {\rm cm}^{-2}$ at $R = 7$~kpc) so it could have a
radius for {\Mg} absorption of $\sim$25~kpc (\cite{bow95b}), thus
covering as much as $\sim 25$\% of the halo (not including the stream!).  
The effect of accounting for satellite absorbers in our models would 
yield an increase in predicted numbers of absorbing galaxies at all
impact parameters.
Again, the implication, if we are to obtain 58 absorbers in 51 fields, is 
that the covering factor of the {\it halo proper}\/ around galaxies 
would be reduced well below $\sim70$\%.
This would represent a significant shift in the design of our
models.

It is highly unlikely that a single pre--dominant geometric cross--section
can be invoked to predict the presence or non--presence of absorption.
For one, it is not clear to what degree merging history plays
a role in geometrically distributing {\it absorbing}\/ gas.
Similar merging scenarios at different redshifts might be expected
to produce different absorption properties, given redshift 
evolution of the meta--galactic UV background and the influence
of ionization on the strength of absorption and the dissipative
properties of the gas.
Also, it is not yet clear at what level disk/halo interactions, such
as galactic fountains (\cite{bregman80}) or chimneys (\cite{mm88}),
contribute to the gas content of extended halos.
Since the gas dissipation rate is density dependent, it is possible
that the statistically observed cross--section of absorbers is a 
function of {\Mg} column density.
Gas that dissipates on a few Gyr timescale would likely infall
(arising from merging, extra--galactic accretion, or disk/halo
processes) onto the disk, contributing to its thickness and affecting
its chemical evolution.
Gas that fails to dissipate, yet maintains $\tau_{LL} \sim 1$ at
a minimum, would likely remain distributed in the halo, its absorption
strength being governed by the galactocentric pressure.
This component of the material may not always remain bound to the
galaxy.
This picture has the prediction that, as seen in high resolution
spectra, low column clouds should exhibit the full range of velocities
across the absorption profile, often detectable in the wings of
stronger lines, or suppressing the continuum between strong lines,
whereas high column clouds should have an internal velocity dispersion 
consistent with a population of randomly oriented disks.
The consequences for inferring
the presence or non--presence of absorption from a known galaxy
with a given impact parameter are not clear.
It may be that the statistical geometric cross--section is
sensitive to the equivalent width threshold {\it and}\/ the epoch 
from which the sample of galaxies is drawn.
An $W_0$ threshold of $0.3$~{\AA} at $0.3 \leq z \leq 1.0$ may sample
a relative contribution from both halos and disks.


\section{Testing Absorber Cross--Section and Kinematics}

There are simple predictable kinematic and absorption
properties that may be useful for discerning between these two
geometries.
Both tests require high spatial resolution imaging of the galaxies
identified as absorbers.
The kinematic tests also require high resolution spectra of the absorption.
Both sets of data will soon be available [(HST images)
Steidel \& Dickinson, private communication;
(HIRES/Keck spectra) Churchill (1996)].

\subsection{Disk Geometry and Associated Kinematics}

Absorption associated with galaxy disks is likely to exhibit
the characteristic signature of rotating galaxies (cf.~\cite{lan92}).
As such, correlations between the strength and variation (spread) in
the observed absorption with increased galaxy inclination provides a
statistical test to discern absorber geometry.
In Fig.~5(a), we show a scatter plot of the distribution of absorber
equivalent widths $W_0$ as a function of galaxy inclination $i$.  
Note that nearly edge--on galaxies are predicted to have a median
$W_0 \sim$ 2--4~{\AA}.
If absorption arises primarily in the disk, the absorption exhibits
a larger mean $W_0$ and larger spread as the galaxy becomes more
inclined.
If absorption is primarily generated in the spherical halos of
galaxies, there is no such correlation predicted from our models.

From HST images, it will also be possible to measure the angle of the line
of sight relative to the position angle of the projected galactic
major axis (the equivalent of our $\theta$).  
In Fig.~5(b) we illustrate that, statistically, if absorption arises
in a disk geometry there should be twice as many absorbers along the
major axis as along the minor axis.
Non--absorbers exhibit the opposite behavior.  
In a simplistic spherical case, there is no relationship between the
clouds and the disk, and thus the relative number of absorbers
to non--absorbers exhibits no dependence on the position angle,
$\theta$.

Establishing statistical kinematic trends will rely on accurate
measurements of the absorption line subcomponents and of the
absorbing galaxy's systemic redshifts.
For our disk models, we have computed the difference between the line of sight
velocity seen in absorption and the galaxy systemic velocity.
For complex absorption profiles with several subcomponents, one can
think of the velocity seen in absorption as an optical depth weighted
mean (\cite{hobbs73}).
We assume this weighted mean of the absorption velocity can, on
average, be represented by the line of sight component of the
Tully--Fisher velocity given by $L_K/L^*_K = [V(L_K)/V(L^*_K)]^{3.2}$ 
(\cite{pie88,for95}).
The difference between the absorption velocity and the
systemic velocity of an $L_K$ galaxy is then
\begin{equation}
\Delta V/V(L_K) = (D/R) \cos{\theta} \sin{i} .
\end{equation}
We illustrate in Fig.~6(a) that large $\Delta V/V(L_K)$ can
result when the galaxy has a relatively large inclination.  
The distribution spread results from the various $\theta$ of the line
of sight (eg.~the small values at large $i$ result when the line of
sight passes close to the minor axis).  
Indeed, as we show in Fig.~6(b), there is a strong statistical
relationship between $\Delta V/V(L_K)$ and $\theta$.
In this case, the spread is a result of various galaxy inclinations.
For establishing these trends observationally, it is important to
keep in mind that the accuracy of galaxy systemic redshifts can
usually be measured with accuracy no greater than $\sim 30$~{\kms},
but even with this uncertainty, the general trends are easily
discernible for the expected $\sim 200$~{\kms} rotation speeds of
galaxies.

\subsection{Spherical Geometry and Infall Kinematics}

A very attractive picture, in which absorbing gas is distributed in
more or less a spherical geometry, is radial infall of intergalactic
material.
Steidel \& Sargent (1992) observed that redshift evolution of the co--moving
cross--section of {\Mg} absorbers depends strongly upon the
equivalent width limit of the sample.
The observed distribution of equivalent widths of {\Mg} absorbers
changes such that the absorption strength becomes weaker with time.
The argument for infall (\cite{ste93b}) is based upon the conclusion
that this evolution is not provided by changes in ionization
conditions or metallicities, since {\Mg} absorption almost always exhibits
saturation.
Since the sizes of the overall absorbing regions apparently do not
evolve [for $W_0(\lambda 2796) \geq 0.3$~{\AA}, the cross--section is
consistent with no evolution], the inference is that either the numbers or
the velocities of the clouds are decreasing with time.
In a spherical halo model, these ideas are further supported by the
expected short lifetimes of $T \sim 10^4$~K clouds moving in a
virialized halo (\cite{mo95}), so that a reservoir for replenishment
of gaseous material is required to account for the non--evolving halo
size over a Hubble time.

If absorption does arise from clouds distributed in a spherical halo
with infall kinematics, the velocity width of the overall absorption
profile should exhibit a trend with impact parameter.
For smooth mass distribution the trend is quite simple, as we illustrate
with the solid line in Fig.~7 for an $L^*_K$ galaxy with $R_h = R^*$.
The total velocity width is produced by the absorbing gas moving with the
extreme line of sight velocities toward and away from the observer.
This width is a maximum at $D=0$ and decreases to no absorption for
$D \geq R_h$.
In the case of individual infalling clouds, the profile velocity width
is due to the maximum velocity difference between the two most
kinematically extreme clouds.
For the case of a constant infall velocity (no dependence upon $R$),
we compute the profile velocity width [normalized by $V(L_K)$]
for model S2, based upon our Monte--Carlo simulations.
We illustrate the results as a scatter diagram in Fig.~7.
In cases of lines of sight that pass through only one absorbing
cloud no point is plotted.
Note that the general trend given by the smooth distribution is
present, though the scatter is quite severe.

The features that are most relevant to future observations
are the deficit of small profile widths at small impact parameter
and of large profile widths at large impact parameter.
For the case of spherical infall, we re--emphasize that such observed
trends would be corroborated if there was no evidence for a dependence
on the galaxy orientation.
However, it is not unlikely that infall would be influenced by the
non--symmetric potential of the galaxy close in (small $D$), even if
it were dominated by dark halos at large $D$.
If so, then the spherical case presented in Fig.~7 would still apply
for large impact parameters, but the small impact parameter trends may
exhibit additional scatter and show some connection to the galaxy
orientation.


\section{Conclusions}

In this paper, we have been motivated by the central question:
what is the predominant geometrical cross--section of galactic gas
which is known to give rise to {\Mg} absorption in the spectra of
QSOs?
It has been our aim to ascertain if a single geometric cross--section
can be inferred as the general shape that governs the presence or
non--presence of absorption in a given galaxy of known properties
intervening to a QSO.
To address this question, we have performed Monte--Carlo simulations
of {\Mg} absorption produced in either a spherical or in a disk
geometry.
The models were designed to recover the observed statistical properties
of {\Mg} absorbers, namely the distribution of $W_0(\lambda 2796)$,
given by $n(W_0)dW_0 = W_0^{-1.65} dW_0$ (\cite{ss92}), the 
scatter in the $W_0 - D$ plane, and a near--unity covering factor.
We assume throughout that each model represents a single population
of absorbers, and used the $D-L_K$ plane and the
Holmberg--like relation (Eq.~1)
to compare our models to currently available observations.
We find that:
 
\begin{enumerate}

\item 
A population of randomly oriented disks can produce a geometric
cross--section comparable to that of spherical halos.
The relative contribution is a function of both the disk
thickness and its extent beyond the halo proper (see Fig.~1).
Additionally, lines of sight that impact highly inclined disks at
large radii likely yield absorption with $W_0 \geq 0.3$~{\AA}, 
due to the increased column density along the line of sight, the
increased velocity dispersion of the clouds in the disk, and 
material that extends even as little as 5~kpc above and below the disk
(such as with warps).

\item 
Both spherical and disk models for which the parameters are tuned to
recover $n(W_0)dW_0$, yield similar fractions of non--absorbers at
impact parameters smaller than the ``boundary'' given by $R(L_K) =
R^*(L_K/L^*_K)^{0.15}$.  Spherical models can be designed to yield
small fractions of absorbers above the $R(L_K)$ boundary, but
disk models predict larger numbers due to lines of sight at large
impact parameters through highly inclined disks.
As a caveat, we note that the largest observed $W_0$ may be sampling
galaxy pairs (\cite{chu96}), which implies that the distribution
derived from single galaxies would have a steeper slope.
The observed trend for {\Mg} absorbers to become less
strong with time (\cite{ss92}) may have some interesting implications
for the questions we have attempted to address in this paper.
One possible view is that a ``pre--dominant geometry'' in which {\Mg}
absorption arises could be changing with time (evolving population),
based upon the notion that numerous and kinematically diverse
optically thick clouds could be associated with halos or merging
events at
redshifts of $z \geq 1$ and could then be evolving into thick and
warped disks at redshift $z \leq 1$ with remnants in the form of high
velocity clouds still accreting at the present epoch.
Such accretion is known to be occurring in the Galaxy
[cf.~the recently discovered dwarf galaxy in Sagittarius,
Ibata, Irwin, \& Gilmore (1994)], whose halo appears
to {\it not}\/ be dynamically mixed (\cite{majewski96}).
Indeed, such mergers are predicted to leave detectable ``streaming
tails'' or ``moving groups'' in the halo for timescales of a Gyr or
more (\cite{kvj95}).

\item 
Our models qualitatively reproduce the observed $W_0 - D$ distribution
(Fig.~3 in S95), which suggests that our scheme to model the ``clumpy
structure'' of {\Mg} absorbing gas in galaxies is appropriate, and
provides some support that the covering factors predicted by our models 
are likely to be fairly accurate.
We find that the small number of observed non--absorbers in the SDP
sample does not {\it necessarily}\/ imply near--unity covering factor 
to the equivalent width limit $W_0(\lambda 2796) = 0.3$~{\AA}, but
can be consistent with covering factors as small as 70--80\%.
In the case of disks we can think of an {\it effective}\/ covering
factor determined by considering a population of randomly oriented disks.
If {\Mg} absorbing galaxies do indeed have covering factors approaching 
those predicted by our models, then the 58 absorbers detected by SDP
in 51 QSO fields imply that the number of galaxies per 10{\arcsec}
QSO field may be as small as $\sim 0.65$ in the redshift regime 
$0.3 \leq z \leq 0.9$ in which case $dN/dz$ is $\sim 1$--$\sigma$
smaller than the observed value.  Spherical cloud
models with larger covering factors cannot be made consistent with the
observed $n(W_0)dW_0$ and $W_0 - D$ distributions and also reproduce 
as {\it few}\/ as 58 absorbers in 51 fields in our``SDP survey''
simulations.

\item
Since our models are either entirely ``halo'' or ``disk'', they both 
likely {\it under predict}\/ the number of observed absorbers.   The
implication is that the halo ``component'' of a model incorporating
both disk and halo would necessarily have a covering factor $\leq
70$\%, if we are to obtain 58 absorbers in our 51 simulated absorber
fields.
In our simulations, we find a gap in the $W_0 - D$ plane at small
$W_0$ for highly inclined disks at small impact parameters ($D \leq
15h^{-1}$~kpc).
We attribute this to the fact that our model disks are not {\it
physically}\/ extended above or below their plane.
Since most or all observed absorbers in this impact regime are DLAs
(S95), and since we infer that these absorbing galaxies indeed have
``thick'' disks based upon our models, we suggest that many of the
small impact DLAs arise from highly inclined thick disks.

\item
Our models, which are an unbiased sample of QSO fields in which all
galaxy redshifts are ``measured'' to a limiting $L_K$, generally yield
no clear boundary between absorbing and non--absorbing galaxies.  
This is consistently true for disk models, but for spherical models the
``distinctness'' of the boundary depends strongly on parameter
choices.  
A paucity of absorbers above the $R(L_K)$ line could be used
to argue against a disk geometry.  
But we note that a relatively larger number of absorbers above the
line could be consistent with either geometry.

\item 
Using the the observed number of non--absorbing galaxies that fall
below the $R(L_K)$ line in S95 as a discriminant between a population
of spherical (halo) absorbers and disk absorbers, we find that, once 
we account for (1) possible selection biases toward absorbing fields,
and, (2) possible mis--identifications of a few galaxies,
we cannot definitively distinguish which geometry governs the presence or
non--presence of absorption.  
We find that the probability for detecting only a few non--absorbers
below the $R(L_K)$ line is not negligible for either our sphere or
disk models.  
The final results of the SDP survey, of which S95 is only a
preliminary report, are likely to provide a demanding test for the
validity our models.
In principle, the number of absorbing galaxies that fall above the
$R(L_K)$ line may be used as a discriminant as well, with small
numbers inconsistent with a disk cross--section but consistent with
certain spherical cloud models.
A complete unbiased sample would allow a more robust measure of
the fraction of absorbing galaxies both {\it above}\/ and {\it
below}\/ the $R(L_K)$ line.

\item 
Upcoming high spatial resolution images (HST) of the absorbing
galaxies and high resolution spectra (few {\kms}) promise to provide
the necessary data to settle the issue of how {\Mg} absorbing gas is
distributed in and around galaxies.
In particular, if absorption does arise in the disks of galaxies, both
the disk inclination and the angle subtended between the location of
the line of sight on the plane of the sky and the galaxy major axis
are expected to be strongly correlated with presence of absorption and
with the velocity difference between the absorption centroid and the galaxy
itself (see Fig.~5).
If absorption arises purely within halo clouds (infalling or otherwise
kinematically distinct from the disk), no such trends with disk
orientation are expected.
No doubt, there will be examples of each, and of both together in
a single system.
The real issue we have attempted to address is which one statistically
governs the prediction of absorption or non--absorption, since this
should yield important clues about the origin and fate of gas for 
theories of galaxy evolution.
For clouds infalling into spherical halos, the width of the overall
absorption profile is expected to decrease with increasing impact
parameter (see Fig.~7), though the scatter can be quite large in the
regime $20 \leq D \leq 40$~$h^{-1}$.
Until these observations and tests are investigated, we submit that
the standard picture of a spherical distribution of halo clouds around
galaxies is not necessarily required by existing data.  
Since, at intermediate redshifts, {\Mg} absorption may be present in
the disks of galaxies out to radii at which {\HI} drops below
$10^{17}$~{\cm2}, we suggest that {\Mg} disks are quite likely to make
a substantial contribution to the population of {\Mg} absorbers.

\end{enumerate}

In the real universe it is likely that both spherical and disk
geometries contribute to the {\Mg} cross--section, and that no single
geometric configuration of discrete absorbing clouds can
identified as being solely responsible for {\Mg} absorbers.
If this is true, then the apparent boundary in the $D-L_K$ plane, which
provides compelling evidence suggestive of a mechanism for
a $K$ luminosity (mass) dependent cut--off in absorbing galaxies,
may indeed result from small number statistics.

As an example of how varied absorption conditions may be,
consider three possible lines of sight through a disk galaxy:
1) A line of sight close ($< 10$--20~kpc) to the center 
of a modestly inclined spiral galaxy must pass through the disk and
may pass through halo clouds or satellite galaxies.  
If the inclination is large, then for absorption to be strong,
the disk would likely have extended material above or below the
plane.
2) If the line of sight passes at larger impact parameter through
a modestly inclined disk galaxy, then the question of the origin
of {\Mg} absorption is more controversial.  It becomes a question of
whether the {\HI} disk extends far enough beyond the optical radius of a
galaxy and/or whether the halo has a significant covering factor of
clouds, high velocity or otherwise.
The relative importance of these two contributions may change with redshift.
3) Very large impact lines of sight will pass too far out in
an inclined disk galaxy for absorption from the disk itself to
dominate.  In these cases, an absorption line may still result
from halo clouds or from satellites or satellite debris and 
would likely be optically thin.
Though there are contributions to absorption profiles
from multiple components in galaxies and their surroundings,
a general geometric cross--section for absorption may still
be definable.  
Chances are that this will be a function of the {\Mg} equivalent width
threshold, or actually, of the column density ``contour''.
Progress on defining the contributions
of halos and disks to these cross--sections is forthcoming
through high--resolution imaging,
while the specific details of what galaxy components
combine to produce absorption will rely on their signatures
in high--resolution spectra.

\acknowledgments

This work was supported in part by NASA grant NAGW--3571 at Penn State
and by the National Science Foundation under Grant No.~PHY94--07194
through the ITP.
CWC would like to acknowledge partial support from the California
Space Institute under a grant issued to S. Vogt.
It is a pleasure to thank our colleagues D. Bowen, S. Charlot, R. Ciardullo,
S. Horner, K. Lanzetta, D. Meyer, C. Norman, P. Shapiro, and J. van Gorkom, 
for their insights.
We are grateful to Chuck Steidel and Mark Dickinson for detailed discussions 
regarding the selection procedures implemented during their {\Mg} absorbing 
galaxy study.



\onecolumn

\include{tables_v07}


\end{document}

%% file: tables_v07.tex

\begin{deluxetable}{lccccccccc}
\tablecaption{Effect of the Variation of Parameters for Sphere Models}
\tablehead
{
& \multicolumn{4}{c}{Model Parameters} & 
\multicolumn{5}{c}{Absorption Properties} \\
\cline{2-5} \cline{6-10} 
\colhead{Model} & 
\colhead{$\alpha$} &
\colhead{$\sigma$} &
\colhead{$N_{\rm c}$} &
\colhead{$W_{\rm min}$ [\AA]} &
\colhead{$R^{*}$ [kpc]} &
\colhead{$f_{\rm abs}$} &
\colhead{$f_{\rm below}$} &
\colhead{$f_{\rm above}$} &
\colhead{$\delta$} \\
\colhead{(1)} &
\colhead{(2)} &
\colhead{(3)} &
\colhead{(4)} &
\colhead{(5)} &
\colhead{(6)} &
\colhead{(7)} &
\colhead{(8)} &
\colhead{(9)} &
\colhead{(10)} 
}
\startdata
S1  & 0.0 & 0.2 & 3 & 0.30 & 34 & 0.46 & 0.76 & 0.08 & 2.3 \nl
S2  & 0.5 & 0.2 & 3 & 0.30 & 35 & 0.48 & 0.78 & 0.09 & 1.8 \nl
S3  & 1.0 & 0.2 & 3 & 0.30 & 35 & 0.47 & 0.79 & 0.08 & 1.4 \nl
S4  & 2.0 & 0.2 & 3 & 0.30 & 36 & 0.48 & 0.80 & 0.10 & 1.2 \nl
S5  & 1.0 & 0.4 & 2 & 0.30 & 37 & 0.46 & 0.66 & 0.22 & 2.1 \nl
S6  & 1.0 & 0.4 & 3 & 0.30 & 35 & 0.47 & 0.78 & 0.09 & 1.4 \nl
S7  & 1.0 & 0.4 & 6 & 0.30 & 36 & 0.53 & 0.93 & 0.03 & 0.4 \nl
S8  & 0.0 & 0.4 & 6 & 0.15 & 37 & 0.48 & 0.83 & 0.04 & 3.5 \nl
S9  & 0.5 & 0.4 & 6 & 0.15 & 37 & 0.48 & 0.84 & 0.04 & 2.1 \nl
S10 & 1.0 & 0.4 & 6 & 0.15 & 37 & 0.49 & 0.85 & 0.04 & 1.4 \nl
S11 & 2.0 & 0.4 & 2 & 0.30 & 39 & 0.47 & 0.66 & 0.22 & 1.7 \nl
S12 & 0.8 & 0.4 & 6 & 0.15 & 37 & 0.49 & 0.85 & 0.04 & 1.6 \nl
\tablecomments{
The model designation is given in column 1.
Column 2 lists the index for $W(R) \sim R^{-\alpha}$ for individual clouds.
The variance $\sigma$ around this mean equivalent width $W(R)$ is
listed in column 3.
Column 4 gives the mean number of clouds along a $D=0$ line of sight.
The minimum equivalent width $W_{\rm min}$ of an individual cloud is
listed in column 5.
Column 6 lists the value of $R(W_0 = 0.3~\hbox{\AA})$ for $L_K^*$ galaxy.
Columns 7, 8, and 9 give the fraction of all galaxies in unbiased
fields that produce absorption, the fraction of galaxies below the
boundary that produce absorption, and the fraction of galaxies above
the boundary that produce absorption, respectively.
In column 10 is the best fit slope of an assumed power law to the $W_0$
distribution.
}
\enddata
\end{deluxetable}


\begin{deluxetable}{lcccccccc}
\tablewidth{33pc}
\tablecaption{Effect of the Variation of Parameters for Disk Models}
\tablehead
{
& \multicolumn{3}{c}{Model Parameters} & 
\multicolumn{5}{c}{Absorption Properties} \\
\cline{2-4} \cline{5-9} 
\colhead{Model} & 
\colhead{$\alpha$} &
\colhead{$\sigma$} &
\colhead{$R_{\rm cut}$} &
\colhead{$R^{*}$ [kpc]} &
\colhead{$f_{\rm abs}$} &
\colhead{$f_{\rm below}$} &
\colhead{$f_{\rm above}$} &
\colhead{$\delta$} \\
\colhead{(1)} &
\colhead{(2)} &
\colhead{(3)} &
\colhead{(4)} &
\colhead{(5)} &
\colhead{(6)} &
\colhead{(7)} &
\colhead{(8)} &
\colhead{(9)} 
}
\startdata
D1  & 1.0 & 0.0 & $\infty$ & 38 & 0.91 & 1.00 & 0.79 & 1.5 \nl
D2  & 1.0 & 0.0 & 2.0      & 34 & 0.52 & 0.75 & 0.23 & 1.9 \nl
D3  & 1.0 & 0.0 & 1.5      & 37 & 0.47 & 0.70 & 0.18 & 1.8 \nl
D4  & 0.0 & 0.0 & 2.0      & 27 & 0.50 & 0.69 & 0.26 & 3.0 \nl
D5  & 1.0 & 0.0 & 2.0      & 34 & 0.52 & 0.75 & 0.23 & 1.9 \nl
D6  & 2.0 & 0.0 & 2.0      & 35 & 0.49 & 0.75 & 0.16 & 1.5 \nl
D7  & 1.0 & 0.4 & 2.0      & 38 & 0.54 & 0.71 & 0.33 & 1.7 \nl
D8  & 1.3 & 0.0 & 2.0      & 34 & 0.50 & 0.74 & 0.20 & 1.8 \nl
D9  & 1.1 & 0.2 & 1.5      & 40 & 0.50 & 0.71 & 0.24 & 1.8 \nl
\tablecomments{
The model designation is given in column 1.
Column 2 lists the index for $W(R) \sim R^{-\alpha}$ for individual clouds.
The variance $\sigma$ around this mean equivalent width $W(R)$ is
listed in column 3.
Column 4 gives the cut--off radius in units of $R(W_0 = 0.3~\hbox{\AA})$.
The value of $ R(W_0 = 0.3~\hbox{\AA})$ for an $L_K^*$ galaxy is
listed in column 5.
Columns 6, 7, and 8 give the fraction of all galaxies in unbiased
fields that produce absorption, the fraction of galaxies below the
boundary that produce absorption, and the fraction of galaxies above
the boundary that produce absorption, respectively.
In column 9 is the best fit slope of an assumed power law to the $W_0$
distribution.
}
\enddata
\end{deluxetable}


\begin{deluxetable}{cccccccc}
\tablewidth{33pc}
\tablecaption{Effect of the Variation of Density of Absorbing Galaxies}
\tablehead
{
\multicolumn{2}{c}{\phantom{Hi Jane}} &
\multicolumn{2}{c}{Full Sample of Fields\tablenotemark{a}} & 
\multicolumn{2}{c}{Absorber Fields\tablenotemark{b}} & 
\multicolumn{2}{c}{Control Fields\tablenotemark{c}} \\
\cline{1-2} \cline{3-4} \cline{5-6} \cline{7-8}
\colhead{$N_{\rm field}$} &
\colhead{$F_{\rm abs}$} &
\colhead{${N_{\rm abs}}$} &
\colhead{${N_{\rm non}}$} &
\colhead{${N_{\rm abs}}$} &
\colhead{${N_{\rm non}}$} &
\colhead{${N_{\rm abs}}$} &
\colhead{${N_{\rm non}}$}
}
\startdata
\cutinhead{Model D8}
0.95 & 0.38 & 64 & 36 & 1.26 & 0.48 & 0 & 0.48 \nl
0.80 & 0.33 & 62 & 30 & 1.21 & 0.40 & 0 & 0.40 \nl
0.65 & 0.28 & 60 & 25 & 1.17 & 0.33 & 0 & 0.33 \nl
\cutinhead{Model S12}
0.95 & 0.37 & 64 & 37 & 1.25 & 0.49 & 0 & 0.49 \nl
0.80 & 0.32 & 62 & 31 & 1.21 & 0.41 & 0 & 0.41 \nl
0.65 & 0.27 & 59 & 25 & 1.17 & 0.33 & 0 & 0.33 \nl
\tablecomments
{
$N_{\rm field}$ is the input mean number of of galaxies per field.  
The fraction of fields that have at least one absorber is given by
$F_{\rm abs}$.
For the full sample, the absorber fields, and the control fields, the
${N_{\rm abs}}$ and ${N_{\rm non}}$ denote the mean number of absorbers
and non--absorbers, respectively.
}
\tablenotetext{a}{Total mean numbers in a sample of 51 absorber fields and 25 control fields}
\tablenotetext{b}{Mean numbers in individual absorber fields}
\tablenotetext{c}{Mean numbers in individual non--absorber fields}
\enddata
\end{deluxetable}


\begin{deluxetable}{lcccccccc}
\tablewidth{33pc}
\tablecaption{Expected Numbers of Non--absorbers ``Below the $R(L_K)$ Line''}
\tablehead
{
& \multicolumn{4}{c}{$N_{\rm field} = 0.95$} & 
\multicolumn{4}{c}{$N_{\rm field} = 0.65$} \\
\cline{2-5} \cline{6-9} 
\colhead{Model} &
\colhead{$N_{10\arcsec}$} &
\colhead{$N_{8\arcsec}$} &
\colhead{$P(\le 2)$} &
\colhead{$P(\le 4)$} &
\colhead{$N_{10\arcsec}$} &
\colhead{$N_{8\arcsec}$} &
\colhead{$P(\le 2)$} &
\colhead{$P(\le 4)$} 
}
\startdata
S2  &  9 & 7 & 0.018 & 0.123 &  6 & 5 & 0.109 & 0.402 \nl
S11 & 13 & 12 & 0.000 & 0.007 & 9 & 8 & 0.009 & 0.080 \nl
S12 &  6 &  5 & 0.110 & 0.412 &  4 & 3 & 0.314 & 0.712 \nl
D7  & 12 &  10 & 0.002 & 0.022 &  8 & 7 & 0.030 & 0.170 \nl
D8  & 10 &  9 & 0.006 & 0.052 &  7 & 6 & 0.051 & 0.260 \nl
D9  & 12 & 10 & 0.002 & 0.023 &  8 & 7 & 0.028 & 0.163 \nl
\tablecomments
{
$N_{10\arcsec}$ is the mean number of non--absorbing galaxies below
the line given by Eq.~(1) in the text.
$N_{8\arcsec}$ is the mean number of non--absorbing galaxies below the line
that are within 8\arcsec\ of the QSO line of sight.
$P(\le 2)$ is the probability of observing two or fewer non--absorbing
galaxies below the line, and $P(\le 4)$ is the probability of
observing four or fewer non--absorbing galaxies below the $R(L_K)$
line.  To mimic the selection procedures of the SDP sample, the
probabilities are computed for the number of galaxies within
8\arcsec\ of the QSO line of sight in the 51 absorber fields plus
the number within 10\arcsec\ for the 25 control fields.
}
\enddata
\end{deluxetable}


\begin{deluxetable}{lcccccccc}
\tablewidth{33pc}
\tablecaption{Expected Numbers of Absorbers ``Above the $R(L_K)$ Line''}
\tablehead
{
& \multicolumn{4}{c}{$N_{\rm field} = 0.95$} & 
\multicolumn{4}{c}{$N_{\rm field} = 0.65$} \\
\cline{2-5} \cline{6-9} 
\colhead{Model} &
\colhead{$N_{10\arcsec}$} &
\colhead{$P(\le 4)$} &
\colhead{$N_{8\arcsec}$} &
\colhead{$P(\le 4)$} &
\colhead{$N_{10\arcsec}$} &
\colhead{$P(\le 4)$} &
\colhead{$N_{8\arcsec}$} &
\colhead{$P(\le 4)$} 
}
\startdata
S2  &  6 & 0.322 & 3 & 0.778 & 5 & 0.375 & 3 & 0.822 \nl
S11 & 13 & 0.001 & 6 & 0.233 & 13 & 0.003 & 6 & 0.279 \nl
S12 & 2 & 0.934 & 1 & 0.992 & 2 & 0.953 & 1 & 0.994 \nl
D7  & 18 & 0.000 & 7 & 0.157 &  16 & 0.000 & 7 & 0.199 \nl
D8  & 11 & 0.008 & 5 & 0.410 & 11 & 0.016 & 5 & 0.464 \nl
D9  & 14 & 0.001 & 6 & 0.238 &  13 & 0.002 & 6 & 0.289 \nl
\tablecomments
{
$N_{10\arcsec}$ is the mean number of absorbing galaxies above the line given
by Eq.~(1) in the text.
$N_{8\arcsec}$ is the mean number of absorbing galaxies above the line
that are within 8\arcsec\ of the QSO line of sight.
$P(\le 4)$ is the probability of
observing four or fewer absorbing galaxies above the $R(L_K)$
line.}
\enddata
\end{deluxetable}

%% file: mgii.bbl
\begin{thebibliography}{}

\bibitem[Bahcall \& Spitzer 1969]{bah69} 
        Bahcall, J.L., and Spitzer, L, Jr. 1969, \apj, 156, L63

\bibitem[Bechtold \& Ellingson 1992]{bec92} 
        Bechtold, J., and Ellingson, E. 1992, \apj, 396, 20 

\bibitem[Bergeron \& Boiss\'e 1991]{ber91} 
        Bergeron, J., and Boiss\'e, P. 1991, A\&A, 243, 344

\bibitem[Bergeron \& Stasi\'nska 1986]{ber86} 
        Bergeron, J., and Stasi\'nska, G. 1986, A\&A, 169, 1

\bibitem[Bergeron \etal 1994]{ber94} 
        Bergeron, J. \etal 1994, \apj, 436, 33

\bibitem[Bowen, Blades, \& Pettini 1995a]{bow95a} 
        Bowen, D.V., Blades, J.C., and Pettini, M. 1995a, \apj, 
        in press

\bibitem[Bowen, Blades, \& Pettini 1995b]{bow95b} 
        Bowen, D.V., Blades, J.C., and Pettini, M. 1995b, \apj, 
        in press

\bibitem[Bregman 1980]{bregman80} 
        Bregman, J. 1980, ApJ, 236, 577

\bibitem[Cardelli, Sembach, \& Savage 1995]{car95} 
        Cardelli, J.A., Sembach, K.R., and Savage, B.D. 1995, \apj,
        440, 241

\bibitem[Cen \etal 1994]{cen94} 
        Cen, R., Miralda--Escud\'e, J., Ostriker, J.P., and Rauch,
        M. 1994, \apj, 427, L9

\bibitem[Churchill 1996]{chu96} 
        Churchill, C.W. 1996, UCSC PhD thesis, in preparation

\bibitem[Churchill, Vogt, \& Steidel 1995]{cwceso}
        Churchill, C.W., Vogt, S.S., and Steidel, C.C 1995 in 
        QSO Absorption Lines, ed.~G. Meylan (Springer Verlag :
        Garching), 153 

\bibitem[Corbelli \& Salpeter 1993]{cor93} 
        Corbelli, E., and Salpeter, E.E. 1993, \apj, 419, 104

\bibitem[Corbelli, Schneider, \& Salpeter 1989]{cor89} 
        Corbelli, E., Schneider, S.E., and Salpeter, E.E. 1989, \aj,
        97, 390

\bibitem[Diplas \& Savage 1991]{dip91} 
        Diplas, A., and Savage, B.D. 1991, \apj, 377, 126

\bibitem[Dove \& Shull 1994]{dov94} 
        Dove, J.B., and Shull, J.M. 1994, \apj, 423, 196

\bibitem[Drinkwater, Webster, \& Thomas 1995]{drink95}
        Drinkwater M.J., Webster, R.L., and Thomas, P.A 1995,
        in Quasar Absorption Lines, ed.~G. Meylan, (Garching :
        Springer--Verlag), 165

\bibitem[Ferland 1988]{ferland88} 
        Ferland, G.J. 1988, Ohio State University Department
        Internal Report, 87--001

\bibitem[Forbes \etal 1995]{for95} 
        Forbes, D.A., Phillips, A.C., Koo, D.C., and Illingworth,
        G.D. 1995, \apj, submitted

\bibitem[Hobbs 1973]{hobbs73} 
        Hobbs, L.M. 1973, ApJ, 180, L79

\bibitem[Hoffman \etal 1993]{hof93} 
        Hoffman, G.L., Lu, N.Y., Salpeter, E.E., Farhat, B.,
        Lamphier, C., and Roos, T. 1993, \aj, 106, 39

\bibitem[Holmberg 1975]{hol75} 
        Holmberg, E. 1975, in Stars and Stellar Systems, 9,
        Galaxies and the Universe, ed. A. Sandage, M. Sandage,
        and J. Kristian, (Chicago: University of Chicago Press), 123

\bibitem[Ibata, Gilmore, \& Irwin 1994]{ibata94}
        Ibata, R.A., Irwin, M.J., and Gilmore, G. 1994, Nature, 370,
        194

\bibitem[Irwin 1995]{irw95} 
        Irwin, J.A. 1995, PASP, 107, 715

\bibitem[Johnston, Spergel, \& Hernquist 1995]{kvj95}
        Johnston, K.V, Spergel, D.N, and Hernquist, L. 1995,
        ApJ, 451, 598

\bibitem[Lanzetta \& Bowen 1992]{lan92} 
        Lanzetta, K.M., and Bowen, D.V. 1992, \apj, 391, 48

\bibitem[Lanzetta \& Bowen 1990]{lan90} 
        Lanzetta, K.M., and Bowen, D.V. 1990, \apj, 357, 321

\bibitem[Lu, Savage, \& Sembach 1994]{lu94} 
        Lu, L., Savage, B.D., and Sembach, K.R. 1994, \apj, 426, 563

\bibitem[MacLow \& McCray 1988]{mm88} 
        MacLow, M.M, and McCray, R. 1988, ApJ, 324, 776

\bibitem[Majewski 1996]{majewski96}
        Majewski, S. 1996, in Formation of the Galactic
        Halo... Inside and Out, eds.~A. Sarajedini and R. Zinn,
        (PASP Conference Series), to preparation

\bibitem[Maloney 1992]{mal92} 
        Maloney, P. 1992, \apj, 398, L89

\bibitem[Maloney 1993]{mal93} 
        Maloney, P. 1993, \apj, 414, 57

\bibitem[Mo 1995]{mo95} 
        Mo, H.J. 1995, in Quasar Absorption Lines,
        ed.~G. Meylan, (Garching : Springer--Verlag), 445

\bibitem[Petitjean \& Bergeron 1990]{pet90} 
        Petitjean, P. and Bergeron, J. 1990, A\&A, 231, 309

\bibitem[Pierce \& Tully 1988]{pie88} 
        Pierce, M.J., and Tully, R.B. 1988, \apj, 330, 579

\bibitem[Rauch 1995]{rau95} 
        Rauch, M., and Haehnelt, M.G. 1995, \mnras, submitted

\bibitem[Salpeter 1993]{sal93} 
        Salpeter, E.E. 1993, \aj, 106, 1265

\bibitem[Salpeter 1995]{sal95}
        Salpeter, E.E. 1995, in The Physics of the Interstellar
        Medium and Intergalactic Medium, ed.~A. Ferrara, C. Heiles,
        C. McKee, and P. Shapiro, (PASP Conference Series), in press

\bibitem[Salpeter \& Hoffman 1995]{salhof95} 
        Salpeter, E.E., and Hoffman, G.J. 1995, \apj, 441, 51

\bibitem[Sargent, Steidel, \& Boksenberg 1988]{ssb88} 
        Sargent, W.L.W., Steidel, C.C, and Boksenberg, A., 1988, ApJ,
        334, 22

\bibitem[Sargent, Boksenberg, \& Steidel 1988]{sbs88} 
        Sargent, W.L.W., Boksenberg, A., and Steidel, C.C 1988, \apjs,
        68, 539

\bibitem[Savage 1993]{sav93} 
        Savage, B.D. \etal 1993, \apj, 413, 116

\bibitem[Savage, Lu, \& Sembach 1995]{sav95} 
        Savage, B.D., Lu, L., and Sembach, K.R. 1995, ed.~G. Meylan,
        (Garching : Springer--Verlag), 119

\bibitem[Sembach, Savage, \& Lu 1995]{sem95} 
        Sembach, K.R., Savage, B.D., and Lu, L. 1995, ApJ, 439, 672

\bibitem[Shandarin \etal 1995]{sha95} 
        Shandarin, S.F., Melott, A.L., McDavitt, K., Pauls, J.L.,
        and Tinker, J. 1995, \prl, 75, 7

\bibitem[Steidel 1995]{ste95} 
        Steidel, C.C. 1995, in Quasar Absorption Lines,
        ed.~G. Meylan, (Garching : Springer--Verlag), 139

\bibitem[Steidel 1993a]{ste93a} 
        Steidel, C.C. 1993, in The Environment and Evolution of
        Galaxies, ed. J.M. Shull and H.A. Thronson, (Kluwer:
        Dordrecht), 263

\bibitem[Steidel 1993b]{ste93b}
        Steidel, C.C. 1993, in Galaxy Evolution: The Milky Way
        Perspective, ed.~S. Majewski, (PASP Conference Series), 49,
        227

\bibitem[Steidel, Dickinson, \& Persson 1994]{sdp94} 
        Steidel, C.C., Dickinson, M., and Persson, S.E. 1994, \apj,
        L75

\bibitem[Steidel, Dickinson, \& Persson 1996]{sdp96} 
        Steidel, C.C., Dickinson, M., and Persson, S.E. 1996, in
        preparation

\bibitem[Steidel \& Sargent 1992]{ss92} 
        Steidel, C.C., and Sargent, W.L.W. 1992, ApJS, 80, 1

\bibitem[Wolfe \etal 1994]{wol94} 
        Wolfe, A.M., Fan, X.--M., Tytler, D., Vogt, S.S., Keane,
        M.J., and Lanzetta, K.M. 1994, \apj, 435, L101

\bibitem[van Gorkom \etal 1993]{van93} 
        van Gorkom J.H. \etal 1993, \aj, 106, 2213

\bibitem[Vogt 1994]{vog94} 
        Vogt, S.S. \etal 1994, SPIE, 2198, 326

\bibitem[Wang 1993]{wang93} 
        Wang, B. 1993, ApJ, 415, 174

\bibitem[York \etal 1986]{york86}
        York, D., Dopita, M., Green, R., and Bechtold, J. 1986, 
        ApJ, 311, 610

\bibitem[Yun, Ho, \& Lo 1994]{yun94}
         Yun, M.S., Ho, P.T.P., and Lo, K.Y. 1994, Nature, 372, 530

\bibitem[Zaritsky \etal 1993]{dennis92}
        Zaritsky, D., Smith, R., Frenk, C., and White, S.D.M. 1993,
        ApJ, 405, 464


\end{thebibliography}
